\documentstyle[preprint,eqsecnum,aps,epsf,floats,tighten]{revtex} 

\begin{document}

\preprint{\tighten \vbox{\hbox{TTP97-31} 
		\hbox{hep-ph/9709457} \hbox{} }}

\title{Calculation of the semileptonic decay width of the $\Lambda_b$ baryon}

\author{Changhao Jin}

\address{Institut f\"ur Theoretische Teilchenphysik, Universit\"at Karlsruhe\\
D--76128 Karlsruhe, Germany}

\maketitle

{\tighten
\begin{abstract}%
We extend the approach based on the light-cone expansion and the heavy quark 
effective theory to the inclusive semileptonic decay of an unpolarized
$b$-flavored hadron. It is applied to calculate the semileptonic decay width
of the $\Lambda_b$ baryon and its ratio to that of the $B$ meson. We obtain
$\Gamma_{\rm SL}(\Lambda_b)= 51\pm 9 \ |V_{cb}|^2\mbox{ps}^{-1}$ and
$\Gamma_{\rm SL}(\Lambda_b)/\Gamma_{\rm SL}(B)= 1.052\pm 0.006$.
From the latter, the semileptonic branching fraction for $\Lambda_b$ is predicted 
to be $(8.7\pm 0.7)\%$. 

\end{abstract}
}

\newpage

\section{Introduction}
Recent measurements \cite{lepb} have shown that the lifetime of 
the $\Lambda_b$ baryon is notably shorter than the lifetime of the $\bar{B^0}$
meson:
\begin{equation}
\frac{\tau(\Lambda_b)}{\tau(\bar{B^0})}=0.79\pm 0.06\, ,
\label{eq:tau}
\end{equation}
suggesting that corrections to the simple spectator decay picture
are significantly different for the $\Lambda_b$ baryon and $B$ mesons.
Studies of the semileptonic branching fractions\footnote{The semileptonic branching
fraction for the $b$-flavored hadron $H_b$ refers to the branching fraction for the
inclusive decay $H_b\to \ell^+\nu$anything, where $\ell$ indicates $e$ or $\mu$
mode.}
for the different $b$-flavored 
hadrons
are also useful to probe the decay dynamics. The semileptonic branching
fraction for $B$ mesons has been measured at the
$\Upsilon(4S)$ resonance to be \cite{PDG}
\begin{equation}
BR_{\rm SL}(B)=(10.43\pm 0.24)\%\, ,
\label{eq:br}
\end{equation}
which is an average over the $\bar{B^0}$ and $B^-$.
Recently, the first measurement of the ratio $R_{\Lambda\ell}=
BR(b-\mbox{baryon}\to\Lambda\ell
X)/BR(b-\mbox{baryon}\to\Lambda X)$ was reported by the OPAL collaboration.
They measure \cite{OPAL} 
\begin{equation}
R_{\Lambda\ell}=(7.0\pm 1.2\pm 0.7)\% ,
\label{eq:R}
\end{equation}
which should be a good approximation to the average $b$ baryon semileptonic
branching fraction. 
These measurements have shown an interesting difference 
in the $\Lambda_b$ and $B$ semileptonic branching fractions.

In this paper we calculate the semileptonic decay width of the $\Lambda_b$
baryon and its ratio relative to that of the $B$ meson\footnote{In the text,
the notation $B$ refers to $\bar{B^0}$ and $B^-$ mesons.}.
The last value can then be converted into a value for 
the semileptonic branching fraction for the $\Lambda_b$ baryon
by using the measured $B$ and $\Lambda_b$ lifetimes and 
the semileptonic branching fraction for $B$ mesons. For this purpose we
must account for the nonperturbative QCD effects in the underlying
weak decays.
 
An approach based on the light-cone expansion
and the heavy quark effective theory has been developed [4 -- 6]
to incorporate nonperturbative QCD effects
in inclusive semileptonic $B$ meson decays.
This approach provides a foundation for
the parton model \cite{parton} for inclusive B decays and furthermore improves it
by including QCD corrections in a systematic way.
This approach accounts correctly for the phase-space effects and produces
a smooth electron spectrum, which is consistent with the experimental
data \cite{jp1}. The calculation of the semileptonic decay width for the
$B$ meson has shown \cite{jin} that the kinematically enhanced nonperturbative
contributions play numerically an important role.

In this paper we extend the
results of [4 -- 6] and compute the inclusive semileptonic decay rate for an unpolarized 
hadron $H_b$ containing a $b$ quark.
The heavy hadron decays involve two large scales: the heavy hadron
mass at the hadron level and the heavy quark mass at the parton level,
which are much greater than the QCD scale. 
Both of them are useful for circumventing nonperturbative QCD effects.
Because of the heaviness of the decaying hadron,
extended regions of phase space in the inclusive semileptonic decays of heavy hadrons 
involve large momentum transfer squared. Therefore, the decay dynamics
is dominated by the light-cone distance. The light-cone dominance attributes 
the nonperturbative QCD effects to a single distribution function. 
On the other hand, the mass of the heavy quark provides
a large limit to construct an effective theory describing the heavy quark interacting
with the gluons in the heavy hadron. This so-called heavy quark effective 
theory (HQET) \cite{hqet} has new (approximate) symmetries that were not manifest in 
the QCD Lagrangian and 
sets a framework for parametrizing nonperturbative effects,
which relates various phenomena (e.g., the spectroscopy and weak decays of hadrons
containing a single heavy quark) to a commen set of parameters, so that 
it has great predictive power.  
Sum rules for the distribution function can be derived by virtue of the operator product
expansion and the HQET method [9 -- 11],
which constrain
the shape of it. Consequently, we can account for the nonperturbative QCD effects
with theoretical uncertainties under control.

This paper is organized in the following manner. We describe in Sec.~II the formalism
showing how the light-cone dominance attributes the nonperturbative QCD effects on the
inclusive semileptonic decays to 
a single distribution function. The semilpetonic decay width for an unpolarized
$H_b$ hadron is expressed in terms of the distribution function. The properties of
the distribution function is discussed in Sec.~III. Two sum rules for the distribution
function are obtained using the operator product expansion and the heavy quark
effective theory. The general formulas for the inclusive semileptonic decay
of the $b$-flavored hadron are applied in Sec.~IV to the $\Lambda_b$ baryon.
Section V contains a conclusion.
 
\section{Formalism}
Consider the inclusive semileptonic decay of an unpolarized hadron $H_b$
containing a $b$ quark
\begin{equation}
H_b\to X_q\ell\bar{\nu_\ell} ,\ \ell=e \ \mbox{or}\ \mu ,
\label{eq:pro1}
\end{equation}
where $X_q$ is any possible hadronic final state containing a charm quark
($q=c$) or an up quark ($q=u$). The decay is induced by weak interactions
and the decay rate is given by
\begin{equation}
d\Gamma= \frac{G_{F}^2\left|V_{qb}\right|^2}{(2\pi)^5P_0}L^{\mu\nu}
W_{\mu\nu}\frac{d^3k_\ell}{2E_{\ell}}\frac{d^3k_{\nu}}{2E_{\nu}} ,
\label{eq:kga}
\end{equation}
where $P$ denotes the momentum of the hadron $H_b$ and $k_{\ell(\nu)}$ and
$E_{\ell(\nu)}$ are the momentum and energy of the electron (antineutrino),
respectively. $V_{qb}$ are the elements of the 
Cabibbo-Kabayashi-Maskawa (CKM) matrix.
$L_{\mu\nu}$ is the leptonic tensor, which results from the summation
over the spins of the charged lepton and antineutrino
\begin{equation}
L^{\mu\nu}= 2(k^{\mu}_{\ell}k^{\nu}_{\nu}+k^{\mu}_{\nu}k^{\nu}_{\ell}-
g^{\mu\nu}k_{\ell}\cdot k_{\nu}+i\varepsilon^{\mu\nu}\hspace{0.06cm}_{\alpha\beta}
k^{\alpha}_{\ell}k^{\beta}_{\nu}).
\label{eq:klepton}
\end{equation}
$W_{\mu\nu}$ is the hadronic tensor, which incorporates all the messy complexity 
of nonperturbative QCD effects for the inclusive process. It 
can be expressed in terms of a current
commutator between the $H_b$ hadron states:
\begin{equation}
W_{\mu\nu}= -\frac{1}{2\pi}\int d^4y e^{iq\cdot y}\frac{1}{2J+1}\sum_s
\langle H_b(P,s)\left|[j_{\mu}(y),j^{\dagger}_{\nu}(0)]\right|H_b(P,s)\rangle ,
\label{eq:comm2}
\end{equation}
where $q$ stands for the momentum transfer to the lepton pair, $q=k_\ell+k_\nu$,
and $j_{\mu}(y) = \bar{q}(y)\gamma_{\mu}(1-\gamma_5)b(y)$ is the weak current.
The spin-J hadron state $|H_b(P,s)\rangle$ satisfies the standard covariant
normalization $\langle H_b(P,s)|H_b(P,s)\rangle=2P_0(2\pi)^3\delta^3({\bf 0})$.  
In general, the hadronic tensor can be decomposed in terms of scalars $W_a(q^2,\, 
q\cdot P)$, $a = 1,\ldots, 5$, as follows :
\begin{equation}
W_{\mu\nu} = -g_{\mu\nu}W_1 + \frac{P_{\mu}P_{\nu}}{M^2} W_2 
 -i\varepsilon_{\mu\nu\alpha\beta} \frac{P^{\alpha}q^{\beta}}{M^2}W_3
         + \frac{q_{\mu}q_{\nu}}{M^2} W_4 
 + \frac{P_{\mu}q_{\nu}+q_{\mu}P_{\nu}}{M^2} W_5 \, ,
\label{eq:exp2}
\end{equation}
where $M$ is the mass of the hadron $H_b$.        

We can express the decay rates in terms of the five hadronic
structure functions $W_a,\, a=1,\ldots, 5$.  The differential decay rate
for the process (\ref{eq:pro1}) in the rest frame of the $H_b$ hadron is
\begin{equation}
\frac{d^3\Gamma}{dE_\ell dq^2dq_0} = \frac{G_F^2|V_{qb}|^2}{16\pi^3 M}
  \left[W_1q^2+W_2(2E_\ell q_0-2E_{\ell}^2-\frac{q^2}{2})+W_3 \frac{q^2}{M}
    (q_0-2E_\ell) \right] \, .
\label{eq:triple2}
\end{equation}
The structure functions $W_4$ and $W_5$ do not appear
above because their contribution is proportional to the square of
the charged-lepton mass and we ignore the lepton masses.

Now we employ the light-cone dominance to simplify the expression for the
hadronic tensor. We proceed along the lines of [4 -- 6].
It is well known  
that integrals like the one in Eq.~(\ref{eq:comm2}) 
are dominated by distances where
\begin{equation}
0 \leq y^2 \leq \frac{1}{q^2}\,.
\end{equation}
For inclusive semileptonic decays (\ref{eq:pro1}), 
$q^2$ is timelike and varies
in the physical range 
\begin{equation}
0 \leq q^2 \leq (M - M_{X_{min}})^2\, ,
\end{equation}
where $M_{X_{min}}$ is the minimum value of the invariant mass of the hadronic 
final state. Because of the heaviness of the hadron $H_b$,
for extended regions of phase space the momentum transfer squared 
is much larger than the QCD scale $\Lambda_{\rm QCD}$.
Therefore, the integral of Eq.~(\ref{eq:comm2}) is dominated by the light-cone
distances in the space-time structure. This allows to replace the commutator
of the two currents with its singularity on the light cone times an opreator
bilocal in the $b$ quark fields, whose on-light-cone matrix element 
between $H_b$ states is also dominant.

The light-cone dominance leads to the expression of the hadronic tensor
in terms of a distribution function 
\begin{equation}
W_{\mu\nu} = 4(S_{\mu\alpha\nu\beta}-i\varepsilon_{\mu\alpha\nu\beta})
 \int d\xi f_{H_b}(\xi) \varepsilon (\xi P_0-q_0) \delta
  \left[ (\xi P-q)^2 -m_q^2 \right] (\xi P-q)^{\alpha} P^{\beta} \, ,
\label{eq:tensor3}
\end{equation}
where $S_{\mu\alpha\nu\beta} = g_{\mu\alpha}g_{\nu\beta} + g_{\mu\beta}
 g_{\nu\alpha} - g_{\mu\nu}g_{\alpha\beta}$ and $m_q$ is the mass of the final
quark $q$. 
The distribution function is defined by
\begin{equation}
f_{H_b}(\xi) = \frac{1}{4\pi M^2}\int d(y\cdot P)e^{i\xi y\cdot P}
\frac{1}{2J+1}\sum_s\langle H_b(P,s)|\bar{b}(0)P\!\!\!/(1-\gamma_5)b(y)
|H_b(P,s)\rangle |_{y^2=0}\, .
\label{eq:distr3}
\end{equation}
It should be emphasized that the form of the distribution function is
not identical for different hadrons. The light-cone dominance implies that 
the five hadronic structure functions can be written in terms of
a single distribution function:
\begin{eqnarray}
W_1 & = & 2[f_{H_b}(\xi_+) + f_{H_b}(\xi_-)]\, ,\\
W_2 & = & \frac{8}{\xi_+ -\xi_-}[\xi_+f_{H_b}(\xi_+)-\xi_-f_{H_b}(\xi_-)]\, ,\\
W_3 & = & -\frac{4}{\xi_+-\xi_-} [f_{H_b}(\xi_+) -f_{H_b}(\xi_-)]\, ,\\
W_4 & = & 0\, ,\\
W_5 & = & W_3\, ,
\end{eqnarray}
where
\begin{equation}
\xi_{\pm}=\frac{q\cdot P \pm \sqrt{(q\cdot P)^2-M^2(q^2-m_q^2)}}{M^2}\, .
\label{eq:root3}
\end{equation} 
The differential decay rate Eq.~(\ref{eq:triple2})
becomes
\begin{equation}
\frac{d^3\Gamma}{dE_\ell dq^2 dq_0} =\frac{G_F^2|V_{qb}|^2}{4\pi^3M}\,
  \frac{q_0-E_\ell}{\sqrt{{\bf q}^2+m_q^2}} \left\{ f_{H_b}(\xi_+)(2\xi_+ E_\ell M-q^2)
    -(\xi_+ \to \xi_-) \right\}\, .
\label{eq:triple3}
\end{equation}
Integrating over the phase space, we obtain
the total semileptonic decay width for the unpolarized hadron $H_b$
\begin{equation}
\Gamma=\frac{G_F^2M^5|V_{qb}|^2}{192\pi^3}\int_r^1d\xi_+\, \xi_+^5f_{H_b}
(\xi_+)\Bigg (1-8\frac{r^2}{\xi_+^2}+8\frac{r^6}{\xi_+^6}-\frac{r^8}{\xi_+^8}
-24\frac{r^4}{\xi_+^4}\mbox{ln}\frac{r}{\xi_+}\Bigg ),
\label{eq:width}
\end{equation}
where $r=m_q/M$ and we have ignored the $f_{H_b}(\xi_-)$ term whose contribution
is negligibly small. Notice that the physical hadron mass $M$ instead of the $b$
quark mass enters Eq.~(\ref{eq:width}), which enhances the decay width with
respect to the free quark decay \cite{jin}.

\section{Properties of the Distribution Function}
Several important properties of the distribution function can be derived from 
field theory.
First of all, because of the conservation of the $b$ quantum number,
the distribution function
is normalized to unity:
\begin{equation}
\int d\xi f_{H_b}(\xi)=\frac{1}{2M^2}P^\mu\frac{1}{2J+1}\sum_s\langle H_b(P,s)\left|\bar b(0)
\gamma_\mu
(1-\gamma_5)b(0)\right| H_b(P,s)\rangle = 1 .
\end{equation}

Consider next $f_{H_b}(\xi)$ in the rest frame of the $H_b$ hadron. In this frame,
\begin{equation}
f_{H_b}(\xi)=\frac{1}{2\pi}\int dy_0e^{iM\xi y_0}\frac{1}{2J+1}\sum_s
\langle H_b(P,s)|b^\dagger (0)P_Lb(y_0)|H_b(P,s)\rangle ,
\end{equation}
where the left-handed projection operator $P_L=(1-\gamma_5)/2$. 
Inserting a complete set of hadronic states between quark
fields and translating the $y_0$ dependence out of these fields, one gets
\begin{equation}
f_{H_b}(\xi)=\sum_m \delta(M-\xi M-p_m^0)\frac{1}{2J+1}\sum_s 
\left| \langle m\left| b_L(0)\right|
H_b(P,s)\rangle \right|^2 ,
\label{eq:ppos}
\end{equation}
where $b_L=P_Lb$. Therefore $f_{H_b}(\xi)$ obeys positivity. The hadronic state 
$|m\rangle$ with momentum $p_m$ is 
physical and must have $0\leq p_m^0\leq M$, thus $f_{H_b}(\xi)=0$
for $\xi \leq 0$ or $\xi \geq 1$. 
Therefore, the support of the distribution function reads
$0\leq \xi \leq 1$.
These results, deduced in the $H_b$ rest frame, 
hold in arbitrary frame due to Lorentz invariance.
Furthermore, Eq.~(\ref{eq:ppos}) gives a probabilistic interpretation of the
distribution function, namely
$f_{H_b}(\xi)$ is the probability of finding a $b$ quark
with a momentum $\xi P$ inside the unpolarized hadron $H_b$. 
In the limit $f_{H_b}(\xi)=\delta(\xi-m_b/M)$, the free quark decay is reproduced
[e.g., Eq.~(\ref{eq:width}) reduces to the free quark semileptonic decay width].

Since the $H_b$ hadron contains a single heavy quark, i.e.~the $b$ quark, the
heavy quark effective theory can be applied. The HQET is successful in describing
various nature of heavy hadrons.
More properties of the distribution function can be deduced exploiting 
the techniques of the operator product expansion and the HQET.

Since the $b$ quark is very heavy within the $H_b$ hadron 
one can extract the large space-time dependence
\begin{equation}
b(y)=e^{-im_bv\cdot y} b_v(y) ,
\label{eq:mlarge}
\end{equation}
where $v$ is the velocity of the initial hadron $H_b$, defined by 
$v=P/M$.
In order to estimate the matrix element of the bilocal operator we must reduce it to
local ones. To this end we make a Taylor expansion of the field in a gauge-covariant form.
This leads to an operator product expansion
\begin{equation}
\bar b(0)\gamma^\beta (1-\gamma_5)b(y)= e^{-im_bv\cdot y}\sum_{n=0}^\infty \frac{(-i)^n}{n!}
y_{\mu_1}\cdots y_{\mu_n}\bar b_v(0)\gamma^\beta (1-\gamma_5)
{\cal S}[k^{\mu_1}\cdots k^{\mu_n}]b_v(0) ,
\label{eq:moper}
\end{equation}
where $k_\mu =iD_\mu=i(\partial_\mu-ig_sA_\mu)$ and
${\cal S}$ denotes a symmetrization.
The Lorentz structure allows to express the matrix element of the local operator 
on the right-hand side of Eq.~(\ref{eq:moper}) between the spin-averaged $H_b$ hadron states
in terms of the $H_b$-hadron momentum:
\begin{eqnarray}
\lefteqn{\frac{1}{2J+1}\sum_s\langle H_b(P,s)|\bar b_v(0)\gamma^\beta (1-\gamma_5)
{\cal S}[k^{\mu_1}\cdots k^{\mu_n}]
b_v(0)|H_b(P,s)\rangle =} \nonumber\\
 & & 2(C_{n0}P^\beta P^{\mu_1}\cdots P^{\mu_n}+
\sum_{i=1}^nM^2C_{ni}g^{\beta\mu_i}P^{\mu_1}\cdots P^{\mu_{i-1}}
P^{\mu_{i+1}}\cdots P^{\mu_n}) \nonumber\\
& &+ \ \mbox{terms \ with \ $g^{\mu_i\mu_j}$} .
\label{eq:mlor}
\end{eqnarray}
The terms with $g^{\mu_i\mu_j}$ can be omitted on the light cone.
Substituting Eqs.~(\ref{eq:moper}) and (\ref{eq:mlor}) into 
Eq.~(\ref{eq:distr3}) yields
\begin{equation}
f_{H_b}(\xi)=\sum_{n=0}^\infty \frac{(-1)^n}{n!}(\sum_{i=0}^n C_{ni})
\delta^{(n)}(\xi-\frac{m_b}{M}) .
\label{eq:mdel}
\end{equation}
Therefore
we obtain the following moment sum rule for the distribution function
\begin{equation}
M_n(m_b/M)=\sum_{i=0}^n C_{ni} ,
\label{eq:msum}
\end{equation}
where the nth-moment about a point $\tilde{\xi}$ of the  
distribution function is in general defined by
\begin{equation}
M_n(\tilde{\xi})=\int_0^1 d\xi  (\xi-\tilde{\xi})^nf_{H_b}(\xi) .
\end{equation}
By definition, $M_0(\tilde{\xi})=C_{00}=1$.

We employ the HQET to estimate further expansion coefficients in 
Eq.(\ref{eq:mlor}). 
In this effective theory the QCD b-quark field $b(y)$ is related to 
its HQET counterpart $h_v(y)$ by means of an expansion in
powers of $1/m_b$,
\begin{equation}
b(y)=e^{-im_bv\cdot y}\Bigg [1+\frac{i/ \mkern -12mu D}{2m_b}+
O(\frac{\Lambda^2_{\rm QCD}}{m_b^{2}})\Bigg ]h_v(y) .
\label{eq:mexp}
\end{equation}
The effective Lagrangian takes the form
\begin{equation}
{\cal L}_{\rm HQET}= \bar h_viv\cdot Dh_v+\bar h_v\frac{(iD)^2}{2m_b}
h_v 
 +\bar h_v\frac{g_sG_{\alpha\beta}\sigma^{\alpha\beta}}{4m_b}
h_v+O(\frac{1}{m_b^2}) ,
\label{eq:mLag}
\end{equation}
where $g_sG^{\alpha\beta}=i[D^\alpha, D^\beta]$ is the gluon field-strength
tensor.
Only the first term in Eq.~(\ref{eq:mLag}) remains in the $m_b\to\infty$
limit, which has the heavy quark spin-flavor symmetry. The other two 
terms give the $1/m_b$ corrections.
Following the method of \cite{manohar}
to relate matrix elements of local operators in full QCD 
to those in the HQET,
the expansion coefficients $C_{ni}$ in Eq.~(\ref{eq:mlor})
can be expressed in terms of the HQET parameters. 
The nonperturbative QCD effects can,
in principle, be calculated in a systematic manner.
In this formalism the moment $M_n(m_b/M)$ is
expected to be of order $(\Lambda_{\rm QCD}/m_b)^n$.
A few resulting coefficients of this method are 
\begin{eqnarray}
C_{10} & = & \frac{5m_b}{3M}E_b(H_b) +
O(\Lambda_{\rm QCD}^3/m_b^3) , \label{eq:mcoef1} \\
C_{11} & = & -\frac{2m_b}{3M}E_b(H_b)+O(\Lambda_{\rm QCD}^3/m_b^3) , \\
C_{20} & = & \frac{2m_b^2}{3M^2}K_b(H_b)+O(\Lambda_{\rm QCD}^3/m_b^3) , \\
C_{21} & = & C_{22} \ = \  0 , \label{eq:mcoef4}
\end{eqnarray}
where the dimensionless HQET parameters
\begin{eqnarray}
K_b(H_b) &=& -\frac{1}{2M}\langle H_b(P,s)|\bar h_v
\frac{(iD)^2}{2m_b^2}h_v|H_b(P,s)\rangle , \label{eq:mK} \\
G_b(H_b) &=& -\frac{1}{2M}\langle H_b(P,s)|\bar h_v\frac{g_sG_{\alpha\beta}
\sigma^{\alpha\beta}}{4m_b^2}h_v|H_b(P,s)\rangle , 
\label{eq:mG}
\end{eqnarray}
and $E_b(H_b)=K_b(H_b)+G_b(H_b)$.
The parameter $K_b(H_b)$ corresponding to the second term in the Lagrangian
Eq.~(\ref{eq:mLag}) measures the kinetic energy of the $b$ quark inside
$H_b$. The parameter $G_b(H_b)$ corresponding to the third term in the Lagrangian
Eq.~(\ref{eq:mLag}) measures the chromomagnetic energy due to the 
spin coupling between the $b$ quark and the light constituents in $H_b$.
Both of them are expected to be of order $(\Lambda_{\rm QCD}/m_b)^2$. 
Thus two sum rules for the distribution function can be derived via the moment 
sum rule (\ref{eq:msum}).
They dertermine
up to order $(\Lambda_{\rm QCD}/m_b)^2$ the mean value $\mu$ and the variance
$\sigma^2$ of the distribution function, which characterize the position of
the maximum and its width, respectively:
\begin{equation}
\mu = \frac{m_b}{M}[1+E_b(H_b)] ,
\label{eq:mmean}
\end{equation}
\begin{equation}
\sigma^2 = \Bigg (\frac{m_b}{M}\Bigg )^2\Bigg [\frac{2K_b(H_b)}{3}-E_b^2(H_b)\Bigg ],
\label{eq:mvar}
\end{equation}
with the definitions
\begin{eqnarray}
& &\mu\equiv M_1(0)=\tilde{\xi}+M_1(\tilde{\xi}) ,
\label{eq:pmean} \\
& &\sigma^2\equiv M_2(\mu)=M_2(\tilde{\xi})-M_1^2(\tilde{\xi}) .
\label{eq:pvar}
\end{eqnarray}
Therefore, the distribution function $f_{H_b}(\xi)$ is sharply peaked
around $\xi =\mu \approx m_b/M$ close to 1 and its width of order 
$\Lambda_{\rm QCD}/M$ is narrow, in agreement with intuitive expectations. 

\section{Application}
As discussed in the last section,
the normalization of the distribution function is fixed to be exactly $1$ by
current conservation 
and the sum rules (\ref{eq:mmean}) and (\ref{eq:mvar}) restrict the position
of the maximum and the width of the distribution function, which improve considerably
the theoretical accuracy, but otherwise its shape is not determined.
In order to calculate the decay rate we shall adopt a distribution function [4 -- 6]
that satisfies the theoretical constraints of the previous section:
\begin{equation}
f(\xi) = N \frac{\xi (1-\xi)}{(\xi -a)^2+b^2}\theta(\xi) \theta (1-\xi)\, ,
\label{eq:ansatz}
\end{equation}
where $N$ is the normalization constant and $a$ and $b$ two parameters,
which depend on the hadron $H_b$ and are related via the sum rules
(\ref{eq:mmean}) and (\ref{eq:mvar}) to the two HQET parameters
$K_b(H_b)$ and $G_b(H_b)$.  
For $a=m_b/M$ and $b=0$, this distribution function reduces to 
a delta function, $\delta(\xi-m_b/M)$, and thus
reproduces the free-quark decay model. 

The semileptonic decay width of the unpolarized hadron $H_b$ can be
calculated by use of Eqs.~(\ref{eq:width}) and (\ref{eq:ansatz}) adding
the known perturbative order $\alpha_s$ correction \cite{pqcd}. 
The semileptonic decay width
of the $B$ meson has been computed \cite{jin} in our approach. A similar calculation
can be done for the $\Lambda_b$ baryon. One needs to know the values of
the corresponding parameters. 

As mentioned in the introduction, the HQET relates various phenomena with each other.
Information on the parameters can be gained with the help of the HQET.
Unlike $B$ mesons, the chromomagnetic contribution vanishes for
the ground-state baryon $\Lambda_b$:
\begin{equation}
G_b(\Lambda_b)=0 ,
\end{equation}
since the light constituents inside $\Lambda_b$ have total spin zero.
The difference between the $b$-quark kinetic energies in the $B$ meson and
in the $\Lambda_b$ baryon can be inferred by using the HQET mass 
formula \cite{rev} .
It follows that $K_b(\Lambda_b)-K_b(B)= 0.0002\pm 0.0006$.
For numerical analyses, we therefore assume the approximate equality:
\begin{equation}
K_b(\Lambda_b)\approx K_b(B)= -\frac{\lambda_1}{2m^2_b}\, ,
\end{equation}
which is further supported by the calculations of the QCD sum rules for the $B$
meson \cite{meson} and for the $\Lambda_b$ baryon \cite{baryon}.
The difference between the $b$ and $c$ quark masses can also be derived 
\cite{rev} from the observed hadron masses by using the mass formula:
\begin{equation}
m_b-m_c=(\overline{M}_B-\overline{M}_D)\Bigg\{1-\frac{\lambda_1}
{2\overline{M}_B\overline{M}_D}
+O(1/m_c^3)\Bigg\} ,
\end{equation}
where the spin-averaged meson masses $\overline{M}_B=\frac{1}{4}(M_B+3M_{B^\ast})
=5.31$ GeV and $\overline{M}_D=\frac{1}{4}(M_D+3M_{D^\ast})=1.97$ GeV. 
Finally, the remaining theoretical input parameters for our calculation are $m_b$,
$\lambda_1$, and the strong coupling constant $\alpha_s$, which is practically
the same set of parameters as in the case of the $B$ meson \cite{jin}. 

For $m_b$ we use
\begin{equation}
m_b=4.9\pm 0.2 \ \mbox{GeV}.
\end{equation}
According to the QCD sum rule calculations \cite{meson,baryon}, we take
\begin{equation}
\lambda_1= -(0.5\pm 0.2) \ \mbox{GeV}^2.
\end{equation}
The strong coupling constant is renormalization scale dependent. 
We vary the scale over the range of $m_b/2\leq \mu_r\leq m_b$ to estimate
the theoretical error due to the scale dependence.
In addition, using the modified parametrization \cite{jin} of the distribution
function with two more parameters $\alpha$ and $\beta$ 
\begin{equation}
f(\xi)=N\frac{\xi (1-\xi)^\alpha}{[(\xi-a)^2+b^2]^\beta}\theta(\xi)
\theta(1-\xi) \ ,
\end{equation}
we find that the values of the semileptonic decay widths for both $\Lambda_b$ and $B$ 
are insensitive to 
the change of the shape of the distribution function if the mean value and 
the variance of it are kept fixed.
This insensitivity diminishes the model dependence.  
We neglect the mass, lifetime, and 
semileptonic branching fraction differences between $\bar{B^0}$ and $B^-$
in all the numerical analyses of this work.
Because of the small $|V_{ub}/V_{cb}|\sim 0.1$, the contributions due to the 
$b\to u$ transition are negligible at the present level of accuracy. The CKM
matrix element $|V_{cb}|$ cancels in the ratio of the $\Lambda_b$ and $B$
semileptonic decay widths making the prediction of the ratio very reliable.
We obtain the following results for the semileptonic decay width of the
$\Lambda_b$ baryon and its ratio to that of the $B$ meson:
\begin{eqnarray}
& &\Gamma_{\rm SL}(\Lambda_b)= 51\pm 9 \ |V_{cb}|^2\mbox{ps}^{-1} ,\\
& &\frac{\Gamma_{\rm SL}(\Lambda_b)}{\Gamma_{\rm SL}(B)}=
1.052\pm 0.006 \, .
\label{eq:wratio}
\end{eqnarray}
The $B$ meson mass $M_B=5.28$ GeV \cite{PDG} and the $\Lambda_b$ baryon mass
$M_{\Lambda_b}=5.62$ GeV from the ALEPH and CDF average \cite{cdf} have been used. 
We find that the semileptonic 
decay width of the $\Lambda_b$ baryon is about $5\%$ larger than the semileptonic
decay width of the
$B$ meson.  Using the theoretical value in Eq.~(\ref{eq:wratio}), togather with
the measured semileptonic branching fraction for the $B$ meson 
in Eq.~(\ref{eq:br})
and the lifetime ratio of $\Lambda_b$ and $B$ in Eq.~(\ref{eq:tau}),
the semileptonic branching fraction for the $\Lambda_b$ is predicted to be
\begin{equation}
BR_{\rm SL}(\Lambda_b)=BR_{\rm SL}(B)\frac{\tau(\Lambda_b)}{\tau(B)}
\frac{\Gamma_{\rm SL}(\Lambda_b)}{\Gamma_{\rm SL}(B)}=(8.7\pm 0.7)\%\, ,
\label{eq:pbr}
\end{equation}
where the error is dominated by the experimental uncertainties on the $\Lambda_b$
and $B$ lifetimes.

\section{Conclusions}
The much more model-independent approach to inclusive semileptonic $B$ meson decays has been 
extended to the inclusive decays $H_b\to X_q\ell\bar{\nu_\ell}$.
We have presented the formula for the semileptonic decay width
of an unpolarized hadron containing a $b$ quark. Eq.~(\ref{eq:width}) shows
that the decay width depends on the mass $M$ of the decaying hadron $H_b$. 
An increase in $M$ amounts to an increase in the semileptonic decay width,
but it does not go as
$M^5$ partly because of the suppression due to the decrease in the mean
value of the distribution function forced simultaneously by the sum rule
(\ref{eq:mmean}).
We applied the formulas to compute the semileptonic decay width of the 
$\Lambda_b$ baryon and found that it is about $5\%$ larger than the semileptonic
decay width of the $B$ meson.   
This suggests that the increase in the total $\Lambda_b$ decay width
observed in the $\Lambda_b$ and $B$ lifetime measurements
originates mainly in the nonleptonic-decay sector.
Our prediction for the semileptonic branching fraction
for $\Lambda_b$ in Eq.~(\ref{eq:pbr}) is compatible with the OPAL result in Eq.~(\ref{eq:R}).
Pure measurements of the semileptonic branching fraction for $\Lambda_b$ are eagerly
awaited.  
Once such measurements become available, the theoretical value for the semileptonic decay 
width of $\Lambda_b$ obtained in this work can be used to gain an independent
determination of the CKM matrix element $|V_{cb}|$.
These measurements and more accurate measurements of the
$\Lambda_b$ and $B$ lifetimes will probe the decay dynamics more extensively and
deeply.
 
\acknowledgments
I am grateful to Bernd Kniehl and the Max-Planck-Institut f\"ur Physik 
(Werner-Heisenberg-Institut) for the warm hospitality, where part of this work was done.

{\tighten

} 

\end{document}